\documentstyle[12pt]{article}
\begin{document}

%\baselineskip=0.3334in
%\begin{flushright} DPUR-84\\  October 1995 \end{flushright}
\null
\vspace{1cm}

\begin{center}
{\Large\bf Quantization of electromagnetic fields\\[.3cm]
in a circular cylindrical cavity}\vspace{1cm}

K. Kakazu\vspace{.1cm}

{\it Department of Physics, University of the Ryukyus,
Okinawa 903-01, Japan}
\vspace{.7cm}

Y. S. Kim\vspace{.1cm}

{\it Department of Physics, University of Maryland, College Park,
Maryland 20742}
\end{center}

\vspace{2cm}

We present a quantization procedure for the electromagnetic field in a
circular cylindrical cavity with perfectly conducting walls, which is based
on the decomposition of the field.  A new decomposition procedure
is proposed; all vector mode functions satisfying the boundary conditions
are obtained with the help of this decomposition.  After expanding the
quantized field in terms of the vector mode functions, it is possible to
derive the Hamiltonian for this quantized system.

\vspace{1cm}
\vfill
\newpage

%--------------------------------------------------------------------
\section{Introduction}
\label{secA}

The behavior of an atom depends on the mode structure of the
electromagnetic field surrounding it \cite{R1}.  The mode structure
in a cavity is determined by its boundary conditions which are quite
different from the periodic boundary conditions used in Lorentz-covariant
electrodynamics.  There are many interesting effects on the interaction
of the atom with the field with various boundary conditions.  It is
possible to calculate them and compare them with experimental data
\cite{R2}.  Thanks to many new ``high-tech'' experimental instruments,
cavity electrodynamics has become one of the most active branches of
physics in recent years.

In quantum electrodynamics, we use the plane waves satisfying the periodic
boundary condition whose quantization volume eventually becomes infinite.
As a result, the orientation of the field is not essential.
On the other hand, the field in a cavity must satisfy the boundary
conditions and the orientation of the field with respect
to cavity walls becomes an important factor.  For this reason, we have to
use vector mode functions to deal with this problem.  These functions
satisfy the boundary conditions and the boundary effect in turn
is contained in the functions.  Thus, for a given cavity, the
most important step in the quantization process is to construct the vector
mode functions.

Indeed, several authors have investigated the quantization procedure for
the field in cavities \cite{R3}.  They have studied the field quantization,
the emission rate, and the atomic level shifts in the cavities with one or
two infinite plane mirrors.  Cavity electrodynamics with rectangular
coordinate system does not present further significant mathematical
problems.  However, the case is quite different for cavities with curved
boundary conditions.  Even for circular cylindrical or spherical cavities,
the quantization procedure has not yet been firmly established.  The purpose
of the present paper is to discuss in detail the mathematical problems in
constructing the vector mode functions with a circular cylindrical boundary
condition.  We hope to discuss the spherical cavity in a later paper.

Recently, we have carried out the field quantization for several different
rectangular boundary conditions including a rectangular tube using the
vector mode functions \cite{R4}.  These functions have been derived
with the help of an orthogonal matrix.  However, the procedure developed
there is not applicable to other cavities in a straightforward manner.

In this paper, we present a more general method of obtaining the vector
mode functions and apply it to the field quantization in a circular
cylindrical cavity with perfectly conducting walls.  This will be
summarized in two theorems.

With this in mind, we first decompose the field into three parts in
the circular cylindrical coordinates in the following section.  Then,
in Sec.\ \ref{secC}, we apply it to the circular cylindrical cavity.  After
obtaining all vector mode functions by using the decomposition, we
arrive at the quantized field and the quantized electromagnetic
Hamiltonian in Sec.\ \ref{secD}.

%------------------------------------------------------------
\section{Decomposition of Electromagnetic fields}
\label{secB}
\setcounter{equation}{0}

Let us now derive the decomposition formula for the electromagnetic
field from Maxwell's equations.  Using the cylindrical coordinate
system, we shall decompose the solutions of Maxwell's equations into
three terms, in preparation for the field quantization within the
circular cylindrical cavity in Sec.\ \ref{secC}.  We emphasize here
again that this decomposition process is the key to the quantization.

Maxwell's equations for the electric field $\bf E$ and the magnetic field
$\bf B$ in free space are given by
\begin{eqnarray}
  & &\nabla\cdot {\bf E} = 0, \label{A1a}\\
  & &\nabla\cdot {\bf B} = 0, \label{A1aa}\\
  & &\nabla\times{\bf E} + \partial_t{\bf B} = 0, \label{A1b}\\
  & &\nabla\times{\bf B} - \frac{1}{c^2}\partial_t{\bf E} = 0,
                                          \label{A1d}
\end{eqnarray}
where $c$ is the velocity of light in free space and $\partial_t =
\partial/\partial t$.  As is well known, it follows from Maxwell's equations
that the fields ${\bf E}$ and ${\bf B}$ satisfy the wave
equations:
\begin{equation}
 \left(\triangle - \frac{1}{c^2}\partial^2_t \right){\bf E}=0, \quad
 \left(\triangle - \frac{1}{c^2}\partial^2_t \right){\bf B}=0,
                                                      \label{A2}
\end{equation}
where $\triangle$ is the Laplacian operator.

The electromagnetic field $\bf E$ and ${\bf B}$ can be written in the
cylindrical coordinates $(r,\varphi,z)$ as
\begin{equation}
  {\bf E} = {\bf E}_T + {\bf E}_z, \quad
  {\bf B} = {\bf B}_T + {\bf B}_z,       \label{A11}
\end{equation}
where ${\bf E}_T = {\bf e}_rE_r + {\bf e}_{\varphi}E_{\varphi}$ is the
transverse component of the field and ${\bf E}_z = {\bf e}_zE_z$.  Here
${\bf e}_r$, ${\bf e}_{\varphi}$, and ${\bf e}_z$ are
the unit vectors in the $r$, $\varphi$, and $z$ directions, respectively.
Similarly, $\nabla \times{\bf E}$, $\nabla\cdot{\bf E}$, and $\nabla\phi$
($\phi$ a scaler function) can also be divided into two parts.  For example,
$\nabla \times {\bf E} = \nabla_T \times {\bf E} + \nabla_z \times {\bf E}$,
where the first term contains the derivatives with respect to $r$ and
$\varphi$, while the second term contains the derivative with respect to
$z$.  For simplicity, the derivatives with respect to $r, \varphi$, and $z$
are described as $\partial/\partial r=\partial_r$, $\partial/\partial\varphi
= \partial_{\varphi}$ and $\partial/\partial z = \partial_z$, respectively.

Since $\nabla_z\times{\bf E}_z = 0$, Eq.\ (\ref{A1b}) gives
\begin{equation}
  \partial_t{\bf B}_z = - \nabla_T \times {\bf E}_T, \quad
  \partial_t{\bf B}_T = - \nabla_T \times {\bf E}_z
       - \nabla_z\times {\bf E}_T.   \label{A12}
\end{equation}
Similarly, from Eq.\ (\ref{A1d}), we have
\begin{equation}
 \frac{1}{c^2}\partial_t{\bf E}_z = \nabla_T \times {\bf B}_T, \quad
 \frac{1}{c^2}\partial_t{\bf E}_T = \nabla_T \times {\bf B}_z
       + \nabla_z\times {\bf B}_T.   \label{A13}
\end{equation}
Equations (\ref{A12}) and (\ref{A13}) give
\begin{eqnarray}
  \frac{1}{c^2}\partial_t^2 {\bf E}_T
   &=& \nabla_T\times \partial_t{\bf B}_z
     - \nabla_z\times(\nabla_T\times {\bf E}_z + \nabla_z\times {\bf E}_T),
                               \nonumber\\
  \frac{1}{c^2}\partial_t^2 {\bf B}_T
   &=& - \frac{1}{c^2}\nabla_T\times \partial_t{\bf E}_z
     - \nabla_z\times (\nabla_T \times {\bf B}_z
       + \nabla_z\times {\bf B}_T),
                               \label{A14}
\end{eqnarray}
which leads to
\begin{eqnarray}
& &\frac{1}{c^2}\partial_t^2{\bf E}_T
    + \nabla_z\times\nabla_z\times{\bf E}_T
    - \nabla_T\times \nabla_T\times{\bf E}_z
   = \nabla\times\partial_t{\bf B}_z
    - \nabla\times\nabla\times{\bf E}_z, \nonumber\\
& &\frac{1}{c^2}\partial_t^2{\bf B}_T
    + \nabla_z\times\nabla_z\times{\bf B}_T
    - \nabla_T\times \nabla_T\times{\bf B}_z
   = - \frac{1}{c^2}\nabla\times\partial_t{\bf E}_z
    - \nabla\times\nabla\times{\bf B}_z,
                            \label{A15}
\end{eqnarray}
where we have used $\nabla_T\times{\bf E}_z = \nabla\times{\bf E}_z$.
Let us note that
\begin{eqnarray}
& &\nabla_z\times\nabla_z\times{\bf E}_T = - \triangle_z{\bf E}_T,
                                         \nonumber\\
& &\nabla_T\times\nabla_T\times{\bf E}_z = - \triangle_T{\bf E}_z,
                                         \nonumber\\
& &\frac{1}{c^2}\partial_t^2{\bf E}_T
   = (\triangle{\bf E})_T = \triangle{\bf E}_T,
                               \label{A16}
\end{eqnarray}
where $(\triangle{\bf E})_T$ is the transverse part of $\triangle{\bf E}$;
the last equation holds in the present coordinates, although it is not
correct in general.  (For example, it is not correct in the spherical
coordinates.)  Then, from Eq.\ (\ref{A15}),
we arrive at
\begin{eqnarray}
& &\triangle_T{\bf E}
    = - \nabla\times\nabla\times{\bf E}_z
     + \nabla\times\partial_t{\bf B}_z,    \nonumber\\
& &\triangle_T{\bf B}
    = - \frac{1}{c^2}\nabla\times\partial_t{\bf E}_z
      - \nabla\times\nabla\times{\bf B}_z.  \label{A17}
\end{eqnarray}

To rewrite Eq.\ ({\ref{A17}), we must obtain an expression for the
components $E_z$ and $B_z$.  Suppose that the field is in a finite region
and expand $E_z$ and $B_z$ in terms of a certain complete system of
functions with mode $s$:
\begin{eqnarray}
  E_z({\bf r},t) &=& \sum_s( E_{zs}({\bf r},t) + {\rm c.c.}),
                                            \nonumber\\
  B_z({\bf r},t) &=& \sum_s(B_{zs}({\bf r},t) + {\rm c.c.}),
                                            \label{Z1}
\end{eqnarray}
where
\begin{equation}
  E_{zs}({\bf r},t) = \tilde E_{zs}({\bf r})e^{-i\omega_{s1}t},
                                   \quad
  B_{zs}({\bf r},t) = \tilde B_{zs}({\bf r})e^{-i\omega_{s2}t}.
                                           \label{Z1-1}
\end{equation}
Here $\omega_{s\sigma}$ $(\omega_{s\sigma}\ge0, \ \sigma=1,2)$ is
determined by using given boundary conditions.  Since $E_z$ and $B_z$
satisfy the wave equation, their components satisfy the Helmholtz
equations:
\begin{equation}
  \triangle E_{zs} = - k_{s1}^2 E_{zs}, \quad
  \triangle B_{zs} = - k_{s2}^2 B_{zs},
                                          \label{Z1-2}
\end{equation}
where
\begin{equation}
  k_{s\sigma}^2 = \frac{\omega^2_{s\sigma}}{c^2}.
                                         \label{Z1-3}
\end{equation}
We assume that the components satisfy
\begin{equation}
  \partial_z^2E_{zs} = - h_{s1}^2 E_{zs}, \quad
  \partial_z^2B_{zs} = - h_{s2}^2 B_{zs},
                                   \label{Z1a}
\end{equation}
where $h_{s\sigma}^2$ is also determined by the boundary conditions.
Then we have two dimensional Helmholtz equations:
\begin{equation}
  \triangle_T E_{zs} = - g_{s1}^2 E_{zs}, \quad
  \triangle_T B_{zs} = - g_{s2}^2 B_{zs},
                                          \label{Z2}
\end{equation}
where
\begin{equation}
  g_{s\sigma}^2 = k^2_{s\sigma} - h^2_{s\sigma}.  \label{Z2a}
\end{equation}

Now we define two functions $F_{\sigma}$ from $E_{zs}$ and $B_{zs}$ with
$g_{s\sigma}^2 \ne 0$ as
\begin{equation}
   F_{\sigma}({\bf r},t)
     = \sum_{g_{s\sigma}^2\ne0} [F_{s\sigma}({\bf r},t)+ {\rm c.c.}]
     = \sum_{g_{s\sigma}^2\ne0} [\tilde{F}_{s\sigma}({\bf r})\,
                e^{-i\omega_{s\sigma}t} + {\rm c.c.}],      \label{Z2b}
\end{equation}
where
\begin{equation}
   F_{s1} = E_{zs}/g_{s1}^2, \quad  F_{s2} = B_{zs}/g_{s2}^2.
                                           \label{Z2b1}
\end{equation}
The functions $F_{\sigma}$ and their components $F_{s\sigma}$ satisfy
the same equations as the $z$ components of the field:
\begin{equation}
  \triangle F_{\sigma}
    = - \frac{1}{c^2}\frac{\partial}{\partial_t} F_{\sigma},
  \qquad
  \triangle_T F_{s\sigma} = - g_{s\sigma}^2 F_{s\sigma}
                                                \label{Z2c}
\end{equation}
{}From Eqs.\ (\ref{Z2b1}) and (\ref{Z2c}) the component $F_{s\sigma}$ is a
solution of the Poisson equation
\begin{equation}
  \triangle_T F_{s1} = - E_{zs}, \quad
  \triangle_T F_{s2} = - B_{zs}             \label{Z2d}
\end{equation}
and then the functions $\triangle_TF_{\sigma}$ satisfy
\begin{eqnarray}
  &&\triangle_TF_1
      = - \sum_{g_{s1}^2\ne0} (g_{s1}^2 F_{s1} + {\rm c.c.})
      = - \sum_{g_{s1}^2\ne0} (E_{zs} + {\rm c.c.}), \nonumber\\
  &&\triangle_TF_2
      = - \sum_{g_{s2}^2\ne0} (g_{s2}^2 F_{s2} + {\rm c.c.})
      = - \sum_{g_{s2}^2\ne0} (B_{zs} + {\rm c.c.}). \label{Z3}
\end{eqnarray}

On the other hand, if there is a component $E_{zs}$ or $B_{zs}$ with
$g^2_{s\sigma} = 0$, Eq.\ (\ref{Z2}) reduces to the two dimensional
Laplace equation: $\triangle_T E_{zs} = 0$ or $\triangle_T B_{zs} = 0$.
Now define $E_{0z}$ and $ B_{0z}$ as
\begin{equation}
   E_{0z} =  \sum_{g_{s1}^2 = 0} (E_{zs} + {\rm c.c.}), \quad
   B_{0z} =  \sum_{g_{s2}^2 = 0} (B_{zs} + {\rm c.c.}), \label{Z3-1}
\end{equation}
which satisfy
\begin{equation}
  \triangle_T E_{0z} = 0, \quad \triangle_T B_{0z} = 0. \label{Z4}
\end{equation}

Then we have a useful expression for $E_z$ and $B_z$; for any $E_z$ and
$B_z$ there exist functions $F_{\sigma}$ given by Eq.\ (\ref{Z2b})
and functions $E_{0z}$ and $B_{0z}$ are given by Eq.\ (\ref{Z3-1})
such that
\begin{equation}
  E_z = - \triangle_T F_1 + E_{0z}, \quad
  B_z = - \triangle_T F_2 + B_{0z}. \label{Z3a}
\end{equation}
It is worth emphasizing again that the functions $F_{\sigma}$ are
constructed with the components $E_{zs}$ and $B_{zs}$ with
$g_{s\sigma}^2\ne0$, while $E_{0z}$ and $B_{0z}$ consist of the
components with $g_{s\sigma}^2 = 0$.

Using Eq.\ (\ref{Z3a}) and defining ${\bf F}_{\sigma}$ by
\begin{equation}
   {\bf F}_{\sigma} = {\bf e}_z F_{\sigma},  \label{Z6a}
\end{equation}
we can rewrite Eq.\ (\ref{A17}) as
\begin{eqnarray}
 & & \triangle_T\Big({\bf E} - \nabla\times\nabla\times{\bf F}_1
     + \nabla\times\partial_t{\bf F}_2\Big) =
     - \nabla\times\nabla\times{\bf E}_{0z}
     + \nabla\times\partial_t{\bf B}_{0z}, \label{A20a}\\
 & & \triangle_T\left({\bf B}
     - \frac{1}{c^2}\nabla\times\partial_t{\bf F}_1
     - \nabla\times\nabla\times{\bf F}_2\right) =
     - \frac{1}{c^2}\nabla\times\partial_t{\bf E}_{0z}
     - \nabla\times\nabla\times{\bf B}_{0z},
                                  \label{A20b}
\end{eqnarray}
where ${\bf E}_{0z}= {\bf e}_zE_{0z}$ and ${\bf B}_{0z} ={\bf e}_zB_{0z}$.
Let us define ${\bf E}_0$ and ${\bf B}_0$ as the quantities in the parentheses
at
the left hand side in Eqs.\ (\ref{A20a}) and (\ref{A20b}), respectively.
The results of this section are then summarized in the following theorem.

\medskip
{\bf Theorem 1:} The field can be decomposed into three components as
follows:
\begin{eqnarray}
 & &{\bf E} = \nabla\times\nabla\times{\bf F}_1
     - \nabla\times\partial_t{\bf F}_2 + {\bf E}_0, \nonumber\\
 & &{\bf B} = \frac{1}{c^2}\nabla\times\partial_t{\bf F}_1
     + \nabla\times\nabla\times{\bf F}_2 + {\bf B}_0,
                                  \label{A21}
\end{eqnarray}
where ${\bf E}_0$ and ${\bf B}_0$ satisfy
\begin{eqnarray}
& &\triangle_T{\bf E}_0
    = - \nabla\times\nabla\times{\bf E}_{0z}
     + \nabla\times\partial_t{\bf B}_{0z},    \nonumber\\
& &\triangle_T{\bf B}_0
    = - \frac{1}{c^2}\nabla\times\partial_t{\bf E}_{0z}
      - \nabla\times\nabla\times{\bf B}_{0z}.  \label{A21a}
\end{eqnarray}
\medskip

Theorem 1 plays a central role in performing field quantization in this
paper.  Take the $z$ component of Eq.\ (\ref{A21}), we find that the $z$
components of ${\bf E}_0$ and ${\bf B}_0$ are given by $E_{0z}$
and $B_{0z}$ in Eq.\ (\ref{Z3a}), respectively.  Furthermore,
Eq.\ (\ref{A21a}) is consistent with Eq.\ (\ref{A17}).

Let us consider the physical meaning of the above decomposition formula.
First it is worth emphasizing that each term in Eq.\ (\ref{A21}) is a
solution to Maxwell's equations.  That is, each term containing
$F_{\sigma}$ ($F_{\sigma}$ term) satisfies Eqs.\ (\ref{A1a}) -
(\ref{A1d}), which is easily shown by using the fact that $F_{\sigma}$
satisfies the wave equation (\ref{Z2c}).  As a result, the third term
${\bf E}_0, {\bf B}_0$ is also a solution to Maxwell's equations, so that
it satisfies Eq.\ (\ref{A17}), as mentioned above.

Next take a particular example of the boundary condition satisfying
$E_{0z}, B_{0z}= 0$. Then the field ${\bf E}_0, {\bf B}_0$ describes
so-called the TEM (transverse electromagnetic) mode \cite{R5}.  Since the
$F_1$ term has no $z$ component of the magnetic field, it becomes
TM (transverse magnetic), while the $F_2$ term becomes TE
(transverse electric), because it does not contain the $z$ component of
the electric field.

In the above particular case where $E_{0z}, B_{0z}= 0$, Maxwell's
equations for the field ${\bf E}_0, {\bf B}_0$ reduce to
\begin{eqnarray}
 & &\nabla_T\cdot{\bf E}_0 = 0, \quad
    \nabla_T\times {\bf E}_0 = 0, \nonumber\\
 & &\nabla_T\cdot{\bf B}_0 = 0, \quad
    \nabla_T\times {\bf B}_0 = 0, \nonumber\\
 & &\nabla_z\times {\bf E}_0 + \partial_t{\bf B}_0 = 0, \nonumber\\
 & &\nabla_z\times {\bf B}_0 - \frac{1}{c^2}\partial_t{\bf E}_0 = 0.
                                   \label{A23}
\end{eqnarray}

%------------------------------------------------------------------------
\section{Determination of Functions $F_{\sigma}$}
\label{secC}
\setcounter{equation}{0}

We are considering here the cavity enclosed by a circular cylindrical wall
with radius $a$ and height $L$: $r<a$, $0<z<L$.  We assume that the cavity
has perfectly conducting walls at $z=0,L$ and at $r=a$.  The tangential
component of the electric field ${\bf E}\vert_{\rm tan}$ and the normal
component of the magnetic field ${\bf B}\vert_{\rm norm}$ must accordingly
vanish at the boundary of the cavity.

The above boundary condition reduces to that for the $z$ components
\begin{eqnarray}
  && E_z = 0, \quad \partial_rB_z = 0,   \qquad (r = a),
                              \label{A24} \\
  && B_z = 0, \quad \partial_zE_z = 0,   \qquad (z = 0,\ L).
                              \label{A25}
\end{eqnarray}
It is easy to get the second condition in Eq.\ (\ref{A24}) if we take
the $\varphi$ component of Eq.\ (\ref{A1d}).  Equation (\ref{A1a}) leads
to the second condition in Eq.\ (\ref{A25}).

Before obtaining the components $E_{zs}$ and $B_{zs}$, we show that the
folowing lemma.

\medskip
{\it Lemma 1:} ${\bf E}_0=0$, ${\bf B}_0 = 0$.
\medskip

{\it Proof:} First we show that $E_{0z}, B_{0z} = 0$.  Consider
the component $E_{0z}$, which satisfies the Laplace equation
$\triangle_T E_{0z} = 0$ [see Eq.\ (\ref{Z4})] with the boundary condition
$E_{0z} = 0$ at the boundary $(r=a)$.  Then we have $E_{0z} = 0$; this is a
well known property of the Laplace equation.

Next consider the component $B_{0z}$ satisfying $\triangle_T B_{0z} = 0$
with the boundary condition $\partial_r B_{0z} = 0$ $(r=a)$.  Then $B_{0z}$
must be a constant, i.e., independent of the variables $r$ and $\varphi$:
$B_{0z} = f(z,t)$.  Applying the two-dimensional Gauss' theorem to
$\nabla\cdot{\bf B}_0 = \nabla_T\cdot{\bf B}_{0T} + \partial_zB_{0z}=0$,
we get $\partial_zB_{0z}=0$, because $B_{0r}=0$ $(r=a)$ and, as a result,
$B_{0z}$ is independent of the variables $r$, $\varphi$, and $z$.
On the other hand, $B_{0z}$ must be zero at $z= 0, L$, which leads to
$B_{0z} = 0$.

As mentioned in the preceding section, the field ${\bf E}_0, {\bf B}_0$
becomes TEM satisfying Eq.\ (\ref{A23}), because $E_{0z}, B_{0z} = 0$.
There exists no TEM in the present cavity.  That is, from Eq.\ (\ref{A23})
we have $\nabla_T\times{\bf B}_0 = 0$ and $\nabla_T\cdot{\bf B}_0 = 0$.
As a result, there exists a function $\varphi$ such that ${\bf B}_0 =
- \nabla_T\varphi$, which gives ${\bf B}_0 = 0$ by the boundary condition
at $r=a$.  Similarly, we get a vector function ${\bf A}$ such that
${\bf E}_0 = \nabla \times {\bf A}$.  The boundary condition leads us to
${\bf E}_0=0$.  Q.E.D.
\medskip

Since ${\bf E}_0, {\bf B}_0 = 0$, Theorem 1 shows that the field is
constructed from the functions $F_{\sigma}$.  Consequently,  from
Eq.\ (\ref{Z2b}), the field only contains the components with
$g_{s\sigma}^2 \ne 0$.  Moreover, it is easy to prove that
$g_{s\sigma}^2 > 0$ by using Gauss' theorem and the boundary condition.

Next we solve the Helmholtz equation (\ref{Z1-2}) for the components
$E_{zs}$ and $B_{zs}$ under the boundary conditions (\ref{A24}) and
(\ref{A25}).  Since $g_{s\sigma}^2>0$, the solution is given by
\begin{eqnarray}
 && E_{zs}({\bf r},t)
      = C_{s1}(t) J_{m}({\chi_{s1}} r/a)\,
        e^{im\varphi} \cos(n\pi z/L),   \nonumber\\
 && B_{zs}({\bf r},t)
      = C_{s2}(t) J_{m}({\chi_{s2}} r/a)\,
        e^{im\varphi} \sin(n\pi z/L),   \label{A26}
\end{eqnarray}
where the mode index is $s=(m,\mu,n)$ $(m=0,\pm1, \pm2, \cdots; \mu=1,2,3,
\cdots; n=0,1,2,\cdots)$, $J_m$ is the Bessel function of the first kind,
$\chi_{s1}\equiv \chi_{m\mu1}$ is the $\mu$th zero point of $J_m$, and
$\chi_{s2}\equiv \chi_{m\mu2}$ is the $\mu$th zero point of $J'_m$, the
derivative of $J_m$.  It follows from Eq.\ (\ref{Z1-1}) that
$C_{s\sigma}(t)\propto \exp(-i\omega_{s\sigma}t)$.

{}From the solution (\ref{A26}) we have
\begin{eqnarray}
  &&\triangle_T E_{zs} = - (k_{s1}^2 - (n\pi/L)^2) E_{zs}
    = - (\chi_{s1}/a)^2 E_{zs}, \nonumber\\
  &&\triangle_T B_{zs} = - (k_{s2}^2 - (n\pi/L)^2) B_{zs}
    = - (\chi_{s2}/a)^2 B_{zs}, \label{A27}
\end{eqnarray}
which gives
\begin{equation}
    g_{s\sigma}^2 = (\chi_{s\sigma}/a)^2, \quad
    k_{s\sigma}^2 = (\chi_{s\sigma}/a)^2 + (n\pi/L)^2.
                                     \label{A28}
\end{equation}

Let us next obtain the functions $F_{\sigma}$, which are defined in
Eqs.\ (\ref{Z2b}) and (\ref{Z2b1}); they are given by
\begin{eqnarray}
  && F_{s1}({\bf r},t) = \frac{a^2 E_{zs}({\bf r},t)}{\chi_{s1}^2}
       \equiv i \sqrt{\frac{\hbar\omega_{s1}}{2\varepsilon_0}}
              a_{s1}(t)\psi_{s1}({\bf r}), \nonumber\\
  && F_{s2}({\bf r},t) = \frac{a^2 B_{zs}({\bf r},t)}{\chi_{s2}^2}
       \equiv i \sqrt{\frac{\hbar\omega_{s2}}{2\varepsilon_0}}
              a_{s2}(t)\psi_{s2}({\bf r}), \label{A30}
\end{eqnarray}
where we have introduced $a_{s\sigma}$ and $\psi_{s\sigma}$:
\begin{eqnarray}
 && a_{s\sigma}(t) = a_{s\sigma}(0) e^{-i\omega_{s\sigma}t},
                                             \label{A30a}\\
 && \psi_{s1}({\bf r},t)
      = c_{s1} J_{m}({\chi_{s1}} r/a)\,
        e^{im\varphi} \cos(n\pi z/L),   \label{A30b}\\
 && \psi_{s2}({\bf r},t)
      = c_{s2} J_{m}({\chi_{s2}} r/a)\,
        e^{im\varphi} \sin(n\pi z/L),   \label{A31}
\end{eqnarray}
where $c_{s\sigma}$ are normalization constants.  The functions
$\psi_{s\sigma}$ have the orthonormality property, which is used in
quantization in the next section.

\medskip
{\it Lemma 2:}
\begin{equation}
  \int_{\rm c} d{\bf r}\, \psi^*_{s\sigma}({\bf r})
  \psi_{s'\sigma}({\bf r})
     = \frac{1}{2}\vert c_{s\sigma}\vert^2 V\alpha_{s\sigma} \delta_{ss'},
                                             \label{A32}
\end{equation}
where $\int_{\rm c}d{\bf r}= \int_{\rm cavity}rdrd\varphi dz$, $V$ is the
cavity volume, and
\begin{eqnarray}
&& \alpha_{s1} \equiv \alpha_{m\mu 1} =
        J_{m+1}^2({\chi_{m\mu 1}}), \nonumber\\
&& \alpha_{s2} \equiv \alpha_{m\mu 2} =
        J_{m}^2({\chi_{m\mu 2}}) -
        J_{m+1}^2({\chi_{m\mu 2}}).
                                             \label{A33}
\end{eqnarray}
Here the quantity $\cos(n\pi z/L)$ in Eq.\ (\ref{A30b}) must
be changed to $1/\sqrt{2}$ when $n=0$.
\medskip

{\it Proof:} The orthonormality property (\ref{A32}) is easily derived
from the following equality for the Bessel functions:
\begin{equation}
 \int_0^a dr\, rJ_{m}({\chi_{m\mu\sigma}} r/a)
 J_{m}({\chi_{m\mu'\sigma}} r/a)
   = \frac{1}{2}\, a^2 \alpha_{s\sigma}\, \delta_{\mu\mu'}.
                                   \label{A34}
\end{equation}

To prove Eq.\ (\ref{A34}), let us first observe the following two
integrals \cite{R6}
\begin{eqnarray}
 \int_0^1 dx\, xJ_m(\chi_ix)J_m(\chi_jx)
 &=& \frac{1}{\chi_i^2-\chi_j^2}
     \Big[\chi_iJ_{m+1}(\chi_i)J_m(\chi_j)
     - \chi_jJ_{m+1}(\chi_j)J_m(\chi_i) \Big]  \nonumber\\
 &=& \frac{1}{\chi_i^2-\chi_j^2}
     \Big[- \chi_iJ'_m(\chi_i)J_m(\chi_j)
     + \chi_jJ_m(\chi_i)J'_m(\chi_j) \Big]
                                   \label{Y1}
\end{eqnarray}
\begin{eqnarray}
 \int_0^1 dx\, xJ_m^2(\chi_ix)
 &=& \frac{1}{2}
     \Big[J^2_m(\chi_i)
     - J_{m-1}(\chi_i)J_{m+1}(\chi_i) \Big]  \nonumber\\
 &=& \frac{1}{2}
     \Big[J^2_m(\chi_i) - J^2_{m+1}(\chi_i)
     - 2J'_m(\chi_i)J_{m+1}(\chi_i) \Big],
                                   \label{Y2}
\end{eqnarray}
where $\chi_i$ $(i = 1, 2, \cdots)$ is a positive real number
and $\chi_i\ne \chi_j$.

Considering that $\chi_{m\mu1}$ is the $\mu$th zero point of
$J_{m}(x)$ and that $J_{m-1}(\chi_{m\mu1})
+ J_{m+1}(\chi_{m\mu1}) = 0$, from Eqs.\ (\ref{Y1}) and (\ref{Y2}),
we have
\begin{equation}
  \int_0^1 dx\, xJ_{m}(\chi_{m\mu1}x)J_{m}(\chi_{m\mu'1}x)
  = \frac{1}{2}\alpha_{m\mu 1}\, \delta_{\mu\mu'},  \label{Y3}
\end{equation}
where $\alpha_{m\mu1}$ is given by Eq.\ (\ref{A33}).
If we change the variable $x$ to $r/a$, Eq.\ (\ref{Y3}) gives
Eq.\ (\ref{A34}) for $\sigma=1$.

Similarly, Eqs.\ (\ref{Y1}) and (\ref{Y2}) lead to
\begin{equation}
  \int_0^1 dx\, xJ_{m}(\chi_{m\mu2}x)J_{m}(\chi_{m\mu'2}x)
  = \frac{1}{2}\alpha_{m\mu2}\, \delta_{\mu\mu'},  \label{Y5}
\end{equation}
where $\alpha_{m\mu2}$ is given by Eq.\ (\ref{A33}).
Equation (\ref{Y5}) gives us Eq.\ (\ref{A34}) for $\sigma=2$.  Q.E.D.

%===============================================================

\section{Vector Mode Functions and Field Quantization}
\label{secD}
\setcounter{equation}{0}

After defining the vector mode functions, we quantize the field and
obtain the quantized Hamiltonian.  The mode functions are constructed
according to the decomposition formula given in Eq.\ (\ref{A21})

In the cavity with the perfectly conducting walls, the decomposition
(\ref{A21}) in Theorem 1 reduces to
\begin{eqnarray}
 &&{\bf E} = \nabla\times\nabla\times{\bf F}_1
     - \nabla\times\partial_t{\bf F}_2, \nonumber\\
 &&{\bf B} = \frac{1}{c^2}\nabla\times\partial_t{\bf F}_1
     + \nabla\times\nabla\times{\bf F}_2,
                                  \label{B1}
\end{eqnarray}
because of ${\bf E}_0, {\bf B}_0 = 0$ [see Lemma 1].

Substituting the functions $F_{\sigma}$ in Eq.\ (\ref{A30}) into ${\bf E}$
in Eq.\ (\ref{B1}), we find
\begin{equation}
 {\bf E}({\bf r},t)
  = i\sum_{s\sigma}\sqrt{\frac{\hbar\omega_{s\sigma}}{2\varepsilon_0}}
    \Big[ a_{s\sigma}(t){\bf u}_{s\sigma}({\bf r})
    - a^*_{s\sigma}(t){\bf u}^*_{s\sigma}({\bf r}) \Big],
                                        \label{B2}
\end{equation}
where the vector mode functions ${\bf u}_{s\sigma}$ are given by
\begin{equation}
 {\bf u}_{s 1} = \nabla\times\nabla\times{\bf e}_z\psi_{s 1}, \quad
 {\bf u}_{s 2} = i\omega_{s2}\nabla\times{\bf e}_z\psi_{s 2}.
                                      \label{B3}
\end{equation}

Similarly, substituting $F_{\sigma}$ into ${\bf B}$ and considering the
equality
\begin{equation}
 \nabla\times{\bf u}_{s1}
    = k_{s1}^2\nabla\times{\bf e}_z\psi_{s1},   \label{B4}
\end{equation}
we obtain
\begin{equation}
  {\bf B}({\bf r},t)
   = \sum_{s\sigma}
     \sqrt{\frac{\hbar}{2\varepsilon_0\omega_{s\sigma}}}
     \Big[ a_{s\sigma}(t)\nabla\times{\bf u}_{s\sigma}({\bf r})
     + a^*_{s\sigma}(t) \nabla\times{\bf u}^*_{s\sigma}({\bf r})\Big].
                                   \label{B5}
\end{equation}

Before obtaining the quantized field and Hamiltonian, let us present some
properties of the mode functions.

\medskip
{\it Lemma 3:}
\begin{eqnarray}
 && \nabla\cdot{\bf u}_{s\sigma} = 0,      \label{B8}\\
 && (\triangle + k_{s\sigma}^2) {\bf u}_{s\sigma} = 0,
                                                 \label{B6}\\
 && {\bf u}_{s\sigma}\vert_{\rm tan} = 0, \quad
    \nabla\times{\bf u}_{s\sigma}\vert_{\rm norm} = 0,
    \quad ({\rm on\ walls}),                 \label{B7}\\
 && \int_{\rm c}d{\bf r}\, {\bf u}^*_{s\sigma}({\bf r})\cdot
     {\bf u}_{s'\sigma'}({\bf r})
        = \delta_{ss'}\, \delta_{\sigma\sigma'}, \label{B9}
\end{eqnarray}

\medskip
{\it Proof:} We give a proof of the orthonormality property (\ref{B9}),
because it is obvious to prove the other equations.  From Eq.\ (\ref{B3}),
the mode function ${\bf u}_{s1}$ is rewritten as
\begin{equation}
 {\bf u}_{s 1}
   = \nabla(\nabla\cdot{\bf e}_z\psi_{s 1})
     - \triangle{\bf e}_z\psi_{s 1}
   = k_{s1}^2{\bf e}_z\psi_{s1} + \nabla(\partial_z\psi_{s1}),
                                 \label{B10}
\end{equation}
which gives
\begin{eqnarray}
 {\bf u}^*_{s1}\cdot {\bf u}_{s'1}
  &=& \Big[ k^2_{s1}k^2_{s'1} - k^2_{s1}h^2_{s'1} \Big]
       \psi^*_{s1}\psi_{s'1}      \nonumber\\
  & & + k^2_{s'1}\partial_z\Big[(\partial_z\psi^*_{s1})\psi_{s'1}\Big]
      + \nabla\Big[ (\partial_z\psi^*_{s1})
        \nabla \partial_z\psi_{s'1} \Big].  \label{B11}
\end{eqnarray}
Taking into account Gauss' theorem and the boundary conditions, we find
that the last two terms in Eq.\ (\ref{B11}) have no effect on the volume
integration of ${\bf u}^*_{s1}\cdot{\bf u}_{s'1}$.  Equation
(\ref{B11}) gives the orthonormality relation for ${\bf u}_{s1}$:
\begin{equation}
     \int_{\rm c}d{\bf r}\, {\bf u}^*_{s1}({\bf r})\cdot
     {\bf u}_{s'1}({\bf r})
        = \delta_{ss'}, \label{B12}
\end{equation}
where we have used Eq.\ (\ref{A32}) and set
\begin{equation}
  c_{s1} = \sqrt{ \frac{2c^2a^2}
          {V\alpha_{s1}\chi^2_{s1}\omega^2_{s1}} }. \label{B13}
\end{equation}
In the case of ${\bf u}_{s2}$, we make use of the equality
\begin{eqnarray}
 \left(\nabla\times{\bf e}_z\psi^*_{s 2}\right) \cdot
 \left(\nabla\times{\bf e}_z\psi_{s' 2}\right)
 &=&{\bf e}_z\psi^*_{s2}\cdot\left(
    \nabla\times\nabla\times{\bf e}_z\psi_{s' 2}\right)
   + \nabla\cdot\Big[{\bf e}_z\psi^*_{s2}\times
     (\nabla\times{\bf e}_z\psi_{s'2})\Big] \nonumber\\
 &=&\frac{\chi^2_{s'2}}{a^2}\, \psi^*_{s2}\psi_{s'2}
     + \nabla\cdot (\psi^*_{s2} \nabla_T\psi_{s'2}).
                                         \label{B14}
\end{eqnarray}
The orthonormality property of ${\bf u}_{s2}$ is then obtained from
Eq.\ (\ref{A32}) if we set the normalization constant:
\begin{equation}
  c_{s2} = \sqrt{ \frac{2a^2}
         {V\alpha_{s2}\chi^2_{s2}\omega^2_{s2}} }.  \label{B15}
\end{equation}
Finally we show that the mode functions ${\bf u}_{s1}$ and ${\bf u}_{s'2}$
are orthogonal to each other.  Since ${\bf u}_{s2}$ has no $z$ component, we
see that
\begin{eqnarray}
 {\bf u}_{s1}^*\cdot {\bf u}_{s'2}
 &=& i\omega_{s'2}\Big(k_{s1}^2{\bf e}_z\psi^*_{s1}
     + \nabla\partial_z\psi^*_{s1}\Big) \cdot
     (\nabla\times{\bf e}_z\psi_{s'2})        \nonumber\\
 &=& i\omega_{s'2}\nabla(\partial_z\psi^*_{s1})\cdot
     (\nabla\times{\bf e}_z\psi_{s'2})        \nonumber\\
 &=& i\omega_{s'2}\nabla\cdot \Big[(\partial_z\psi^*_{s1})
     (\nabla\times{\bf e}_z\psi_{s'2}) \Big],
                                  \label{B16}
\end{eqnarray}
which has no effect on the volume integration of
${\bf u}^*_{s1}\cdot{\bf u}_{s'2}$.  Q.E.D.
\medskip

We are now ready to carry out the field quantization.  To get the quantized
field
the functions $a_{s\sigma}(t)$ are regarded as annihilation operators,
which annihilate photons with indices $s\sigma$ and satisfy the commutation
relation:
\begin{equation}
   [a_{s\sigma}(t), a^{\dagger}_{s'\sigma'}(t)]
      = \delta_{ss'} \delta_{\sigma\sigma'}, \label{B22}
\end{equation}
where $a^{\dagger}_{s'\sigma'}(t)$ are creation operators.  Then the field
given by Eqs.\ (\ref{B2}) and (\ref{B5}) becomes operators.
The quantized electromagnetic Hamiltonian $H_R$ is obtained by using the
equality
\begin{equation}
 \int_{\rm c}d{\bf r}\, (\nabla\times{\bf u}^*_{s\sigma}({\bf r}))
 \cdot(\nabla\times{\bf u}_{s'\sigma'}({\bf r}))
    = k^2_{s'\sigma'}
      \int_{\rm c}d{\bf r}\,{\bf u}^*_{s\sigma}({\bf r})\cdot
      {\bf u}_{s'\sigma'}({\bf r}).
                                     \label{B18}
\end{equation}
These are summarized in the following theorem.

\medskip
{\bf Theorem 2:} The quantized field and the Hamiltonian are given by
\begin{eqnarray}
 &&{\bf E}({\bf r},t)
  = i\sum_{s\sigma}\sqrt{\frac{\hbar\omega_{s\sigma}}{2\varepsilon_0}}
    \Big[ a_{s\sigma}(t){\bf u}_{s\sigma}({\bf r})
    - a^{\dagger}_{s\sigma}(t){\bf u}^*_{s\sigma}({\bf r}) \Big],
                                        \label{B19}\\
 && {\bf B}({\bf r},t)
   = \sum_{s\sigma}
     \sqrt{\frac{\hbar}{2\varepsilon_0\omega_{s\sigma}}}
     \Big[ a_{s\sigma}(t)\nabla\times{\bf u}_{s\sigma}({\bf r})
     + a^{\dagger}_{s\sigma}(t) \nabla\times
       {\bf u}^*_{s\sigma}({\bf r})\Big].
                                       \label{B20}\\
 &&  H_R = \sum_{s\sigma}{1\over2} \hbar\omega_{s\sigma}
         \left(a^{\dagger}_{s\sigma}a_{s\sigma}
         + a_{s\sigma}a^{\dagger}_{s\sigma}\right)
       = \sum_{s\sigma}\hbar\omega_{s\sigma}
         \left(a^{\dagger}_{s\sigma} a_{s\sigma}
         + \frac{1}{2}\right).         \label{B21}
\end{eqnarray}

Theorem 2 will play a fundamental role when we investigate the
electromagnetic property of a physical system in the circular cylindrical
cavity.  There are also important issues in the present cavity including,
for example, spontaneous (stimulated) emission, absorption, the atomic
energy shifts, and the Casimir effect; they will be investigated with the
help of Theorem 2.

It is necessary to expand the field in terms of the vector mode functions
as in Theorem 2.  According to Fourier analysis, the field can be
expanded in terms of, for instance, the periodic functions as in ordinary
quantum electrodynamics.  However, in this case,
the operators $a_{s\sigma}$ and $a^{\dagger}_{s\sigma}$ cannot describe the
physical photons.  Consider the situation where the field
consists of only one mode corresponding to the photon with
$a_{s\sigma}{\bf u}_{s\sigma}$, where ${\bf u}_{s\sigma}$ satisfies the
periodic boundary condition.  This must be impossible, because this field
does not satisfy the boundary conditions; the photon is then fictitious.

%----------------------------------------------------------------------
\section{Conclusions}
\label{secE}
\setcounter{equation}{0}

The quantization procedure of the field in the cavity without
${\bf E}_0, {\bf B}_0$ is as follows: (a) obtain the decomposition
formula (\ref{A21}) in the circular cylindrical coordinates;
(b) solve the Helmholtz equations (\ref{Z1-2}) for the components
$E_{zs}$ and $B_{zs}$ under the boundary conditions (\ref{A24}) and
(\ref{A25});
(c) determine the functions $F_{\sigma}$, which are the solutions of
the Poisson equation (\ref{Z2d}), from $E_{zs}, B_{zs}$;
(d) substitute $F_{\sigma}$ into the decomposition formula (\ref{A21})
and obtain the vector mode functions satisfying the orthonormality
property (\ref{B9});
(e) then we arrive at the quantized field and Hamiltonian in Theorem 2.

As has been shown explicitly in Theorem 2 in Sec.\ \ref{secD}, the energy
$\hbar\omega_{s\sigma}$ of the photon depends on the polarization index
as well as the wave number index.  This is one of the characteristics of
the circular cylindrical cavity.  The photon energy is independent of the
polarization index in a rectangular cavity and in free space.

In the entire process of quantization, the decomposition formula in Theorem 1
plays the essential role.  We must find such a formula in an appropriate
coordinate system if we study other types of cavities.  Also, if
a cavity allows the third term ${\bf E}_0, {\bf B}_0$ in (\ref{A21}),
we have to construct a methodology to deal with the problem.  We hope to
discuss the quantization of electromagnetic fields inside a spherical
cavity in a future publication.

%--------------------------------------------------------------
\section*{Acknowledgments}

We would like to thank Prof.~Y. Ezawa, Dr.~R. Ray, Dr.~G. Gat, Dr.~S.
Kobayashi, Dr. Kurennoy, and Mr.~K. Ohshiro for numerous valuable
discussions.  We are also grateful to Prof. M. S. Kim for bringing some
of the references to our attention.

%----------------------------------------------------------------------

\end{document}